\documentstyle[twoside,fleqn,espcrc2,rotate]{article}

\input{psfig}

\newcommand{\rf}[4]{{\em {#1}} {\bf #2}, #3 (#4)}

\newcommand{\pr}{Phys.\ Rev.\ }

\newcommand{\pl}{Phys.\ Lett.\ }

\newcommand{\np}{Nucl.\ Phys.\ }

\def\be{\begin{equation}}
\def\ee{\end{equation}}
\def\bea{\begin{eqnarray}}
\def\eea{\end{eqnarray}}

\def\qhat{\hat{q}}

\def\half{\frac{1}{2}}
\def\sss{\scriptscriptstyle}
\def\Muv{M_{\sss\rm UV}}
\def\Mir{M_{\sss\rm IR}}
\def\Duv{D_{\sss\rm UV}}
\def\Dir{D_{\sss\rm IR}}
\newcommand{\err}[2]{\raisebox{-0.4ex}
{$\stackrel{\scriptstyle +#1}{\scriptstyle -#2}$}}

\begin{document}

%% Preprint number ADP-98/58-T326
\title{Modelling the gluon propagator\thanks{Talk presented by
D. B. Leinweber}}

\author{D. B. Leinweber\address{CSSM {\em and} 
Department of Physics and Mathematical Physics,\\
The University of Adelaide, SA 5005, Australia}, 
C. Parrinello\address{Department of Mathematical Sciences, University
of Liverpool,\\ Liverpool L69 3BX, England}\thanks{UKQCD
Collaboration}, 
\addtocounter{footnote}{-1}\addtocounter{address}{-2}
J. I. Skullerud\addressmark\footnotemark\addtocounter{address}{-1} 
and A. G. Williams\addressmark}

\begin{abstract}
Scaling of the Landau gauge gluon propagator calculated at $\beta=6.0$
and at $\beta=6.2$ is demonstrated.  A variety of functional forms for
the gluon propagator calculated on a large ($32^3\times 64$) lattice
at $\beta=6.0$ are investigated.
\end{abstract}

\maketitle

\section{Scaling behaviour}

In this note we focus on modelling the gluon propagator calculated in
\cite{jis}.  We use the same conventions and definitions here as in
\cite{jis}.

The lattice gluon propagator is related to the renormalised
continuum propagator $D_R(q;\mu)$ via a renormalisation constant,
\be
D^L(qa) = Z_3(\mu,a) D_R(q;\mu) \, .
\label{eq:renorm-def}
\ee

The asymptotic behaviour of the renormalised gluon propagator in the
continuum is given to one-loop level by
\be
D_R(q) = \frac{1}{q^2}\left(\half\ln(q^2/\Lambda^2)\right)^{-d_D}
\label{model:asymptotic}
\ee
with $d_D=13/44$ in Landau gauge for quenched QCD.

Since the renormalised propagator $D_R(q;\mu)$ is independent of the
lattice spacing, we can use (\ref{eq:renorm-def}) to derive a simple,
$q$-independent expression for the ratio of the unrenormalised
lattice gluon propagators at the same value of $q$:
\be
\frac{D_c(qa_c)}{D_f(qa_f)} = 
\frac{Z_3(\mu,a_c)D_R(q;\mu)/a_c^2}{Z_3(\mu,a_f)D_R(q;\mu)/a_f^2}
= \frac{Z_c}{Z_f}\frac{a_f^2}{a_c^2}
\label{eq:gluon_match_ratio}
\ee
where the subscript $f$ denotes the finer lattice ($\beta=6.2$ in this
study) and the subscript $c$ denotes the coarser lattice
($\beta=6.0$).  We can use this relation to study directly the scaling
properties of the lattice gluon propagator by matching the data for
the two values of $\beta$.  This matching can be performed by
adjusting the values for the ratios $R_Z = Z_f/Z_c$ and $R_a =
a_f/a_c$ until the two sets of data lie on the same curve.
We have implemented this by making a linear interpolation of the
logarithm of the data plotted against the logarithm of the momentum
for both data sets.  In this way the scaling of the momentum is
accounted for by shifting the fine lattice data to the right by an
amount $\Delta_a$ as follows
\be
\ln D_c( \ln(qa_c) ) = \ln D_f( \ln(qa_c) - \Delta_a ) + \Delta_Z
\label{eq:match-data}
\ee
Here $\Delta_Z$ is the amount by which the fine lattice data must be
shifted up to provide the optimal overlap between the two data sets.
The matching of the two data sets has been performed for values of
$\Delta_a$ separated by a step size of 0.001.  $\Delta_Z$ is
determined for each value of $\Delta_a$ considered, and the optimal
combination of shifts is identified by searching for the global
minimum of the $\chi^2/$dof.  The ratios $R_a$ and $R_Z$ are related
to $\Delta_a$ and $\Delta_Z$ by
$R_a = \exp(-\Delta_a)$; $R_Z = R_a^2 \exp(-\Delta_Z)$.

We considered matching the lattice data
using $\qhat=2\pi n/L$ as the momentum variable.  The minimum value for
$\chi^2/$dof of about 1.7 was obtained for $R_a\sim 0.815$.  This value
for $R_a$ is considerably higher than the value of $0.716\pm 0.040$
obtained from an analysis of the static quark potential in \cite{bs}.
From this discrepancy, as well as the relatively high value for
$\chi^2/$dof, we may conclude that the gluon propagator, taken as a
function of $\qhat$, does not exhibit scaling behaviour for the values
of $\beta$ considered here.

\begin{figure}[t]
\begin{center}
\leavevmode
\rotate[l]{\psfig{figure=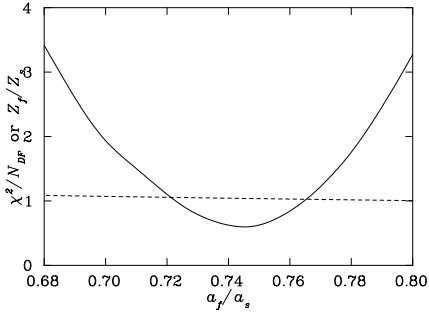,height=2.2in}}
\end{center}
\vspace{-1cm}
\caption{$\chi^2$ per degree of freedom as a function of the ratio of
lattice spacings for matching the small and fine lattice data, using 
$q$ as the momentum variable.  The dashed line indicates the
ratio $R_Z$ of the renormalisation constants.}
\label{fig:match-q}
\end{figure}

Fig.~\ref{fig:match-q} shows the result of
the matching using $q=2\sin(\qhat/2)$ as the momentum
variable.  This gives much more
satisfactory values both for $\chi^2/$dof and for $R_a$.  The minimum
value for $\chi^2/$dof of 0.6 is obtained for $R_a=0.745$.  Taking a
confidence interval where $\chi^2/{\rm dof} < \chi^2_{min} + 1$ gives
us an estimate of $R_a=0.745\err{32}{37}$, which is fully compatible
with the value~\cite{bs} of $0.716\pm 0.040$.

\section{Model fits}
\label{sec:models}

We have demonstrated scaling in our lattice data over the entire range
of $q^2$ considered, and will now proceed with model fits.
The following functional forms have been considered:

\noindent{\bf Gribov\,\cite{gribov}}
\be
D_r(q^2) = \frac{q^2}{q^4+\Mir^4}L(q^2,\Mir)
\label{model-first}
\label{model:lita}
\label{model:gribov}
\ee
{\bf Stingl\,\cite{stingl}}
\be
D_r(q^2) = \frac{q^2}{q^4+2A^2q^2+\Mir^4}L(q^2,\Mir)
\label{model:stingl}
\ee
{\bf Marenzoni\,\cite{mms}}
\be
D_r(q^2) = \frac{1}{(q^2)^{1+\alpha}+\Mir^2}
\label{model:marenzoni}
\ee
{\bf Cornwall\,\cite{cornwall}}
\be
D_r(q^2) =
\left[(q^2+M^2(q^2))\ln\frac{q^2+4M^2(q^2)}{\Lambda^2}\right]^{-1}
\label{model:cornwall}
\ee
{\bf Model A}
\be
D_r(q^2) = \frac{A}{(q^2+\Mir^2)^{1+\alpha}} + \Duv(q^2)
\label{modelA}
\ee
{\bf Model B}
\be
D_r(q^2) = \frac{A}{(q^2)^{1+\alpha}+(\Mir^2)^{1+\alpha}} +
\Duv(q^2)
\label{modelB}
\ee
{\bf Model C}
\be
D_r(q^2) = A e^{-(q^2/\Mir^2)^{\alpha}} + \Duv(q^2)
\label{modelC}
\label{model-last}
\ee
where $D(q^2) \equiv ZD_r(q^2)$ and
\be
\Duv(q^2) = \frac{1}{q^2+\Muv^2}L(q^2,\Muv)
\label{eq:uv-prop}
\ee
\be
L(q^2,M) = \left(\half\ln((q^2+M^2)/M^2)\right)^{-d_D}
\label{eq:log}
\ee
\be
M(q^2) = M\left\{\ln\frac{q^2+4M^2}{\Lambda^2}/
\ln\frac{4M^2}{\Lambda^2}\right\}^{-6/11} \nonumber 
\ee

We have also considered the models A and B, which are constructed from
models A and B by setting $\Muv=\Mir$.  Gribov's and Stingl's models
(\ref{model:gribov}) and (\ref{model:stingl}) are modified in order to
exhibit the asymptotic behaviour of (\ref{model:asymptotic}).
Models~A and B are constructed as generalisations of
(\ref{model:marenzoni}) with the correct dimension and asymptotic
behaviour.

All models are fitted to the large lattice data using the cylindrical
cut defined in \cite{jis,letter}.  The lowest momentum value was
excluded, as the volume dependence of this point could not be
assessed.  In order to balance the sensitivity of the fit between the
high- and low-momentum region, nearby data points within $\Delta(qa) <
0.05$ were averaged.

\begin{figure}[tb]
\begin{center}
\leavevmode
\mbox{\rotate[l]{\psfig{figure=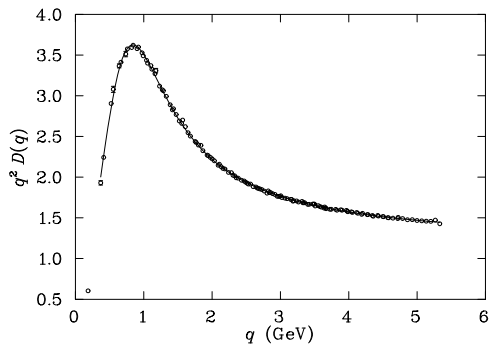,height=2.5in}}}
\end{center}
\vspace{-1cm}
\caption{The gluon propagator multiplied by $q^2$, with nearby points
averaged.  The line illustrates our best fit to the form defined in
(\protect\ref{modelB}).  The scale is taken from the
string tension~\protect\cite{bs}.  }
\label{fig:fit-modelB}
\end{figure}

The $\chi^2$ per degree of freedom and parameter values for fits to all
these models are shown in table~\ref{tab:fit-params}.  It is clear
that model B accounts for the data better than any of the
other models.  The best fit to this model is illustrated in
fig.~\ref{fig:fit-modelB}.

\begin{table*}
\caption{Parameter values for fits to models
(\protect\ref{model-first})--(\protect\ref{model-last}).  The
values quoted are for fits to the entire set of data.  The errors
denote the uncertainties in the last digit(s) of the parameter values
which result from varying the fitting range \protect\cite{next}.}
\label{tab:fit-params}
\begin{tabular*}{\textwidth}{l@{\extracolsep{\fill}}cc@{\extracolsep{0.6cm}}cccc}
\hline
Model & $\chi^2/$dof & $Z$ & $A$ & $\Mir$ & $\Muv$ or $\Lambda$ & $\alpha$
\\ \hline
Gribov & 1972 & 2.19\err{31}{15} & & 0.23\err{14}{1} \\
Stingl & 1998 & 2.2 & 0 & 0.23 \\
Marenzoni & 163 & 2.41\err{0}{12} & & 0.14\err{4}{14}
 & & 0.29\err{6}{2} \\
Cornwall & 50.3 & 6.5\err{7}{9} & & 0.24\err{3}{16} &
0.27\err{7}{7}  \\
A & 2.96 & 1.54\err{10}{20} & 1.24\err{21}{21}
 & 0.46\err{2}{14} & 0.96\err{47}{17} & 1.31\err{16}{43} \\
A' & 3.73 & 1.71\err{9}{0} & 0.84\err{0}{29} & 0.48\err{2}{17}
 & & 1.52\err{12}{37} \\
B & 1.57 & 1.78\err{45}{20} & 0.49\err{17}{6}
 & 0.43\err{5}{1} & 0.20\err{37}{19} & 0.95\err{7}{5} \\
B' & 4.00 & 1.62\err{3}{4} & 0.58\err{5}{1} & 0.40\err{9}{2}
 & & 0.92\err{17}{1} %0.919(+166/-2) 
\\
C & 47 & 2.09\err{30}{12} & 29\err{166}{2}
 & 0.22\err{0}{16} & 0.14\err{11}{10} & 0.49\err{0}{16} \\
\hline
\end{tabular*}
\end{table*}

\section{Discussion and outlook}
\label{sec:discuss}

We have demonstrated that the gluon propagator exhibits scaling over a
wide range of momenta $q$.  The data are consistent with a functional
form $D_r(q^2) = \Dir + \Duv$, where
\be
\Dir = \half \frac{1}{q^4+M^4} ,
\label{eq:ir-prop}
\ee
$M =0.81\err{9}{2}$ GeV, and $\Duv$ is the asymptotic form given by
(\ref{eq:uv-prop}) and (\ref{eq:log}).  A more detailed
analysis~\cite{next} of the asymptotic behaviour reveals that the
one-loop formula (\ref{model:asymptotic}) remains insufficient to
describe data in the region
$q^2\sim 25{\rm GeV}^2$.

Issues for future study include
the effect of Gribov copies and of dynamical fermions.
%% Neither is expected to have a large effect on our main conclusion
We also hope to use improved actions to
perform realistic simulations at larger lattice spacings.  This would
enable us to evaluate the gluon propagator on larger physical volumes,
giving access to lower momentum values.

\section*{Acknowledgments} 

The numerical work was mainly performed on a Cray T3D at EPCC,
University of Edinburgh, using UKQCD Collaboration time under PPARC
Grant GR/K41663.  Financial support from the Australian Research
Council is gratefully acknowledged.

\end{document}